**Growth and structural transitions of core-shell nanorods in nanocrystalline Al-Ni-Y**

Tianjiao Lei [1], Mingjie Xu [1,2], Jungho Shin [3], Daniel S. Gianola [3], Timothy J. Rupert [1,*]

[1] Department of Materials Science and Engineering, University of California, Irvine, CA 92697, USA

[2] Irvine Materials Research Institute, University of California, Irvine, CA 92697, USA

[3] Materials Department, University of California, Santa Barbara, CA 93106, USA

* Email: trupert@uci.edu

**Abstract**

Unique nanorod precipitates with a core-shell structure are found to nucleate from the grain boundaries of a bulk nanocrystalline Al-Ni-Y alloy fabricated via powder consolidation, contributing significantly to stabilization and strengthening. The local structure, chemistry, and evolution of these features during annealing is reported here. In the as-consolidated state, the nanorods can be either structurally ordered or disordered, yet a consistent chemical patterning is found where the core is primarily Al plus C while the shell is enriched with Y. As annealing time increases, more nanorods transform to an ordered structure as they coarsen while the core composition remains unchanged. In contrast, the shell chemistry transitions from Y-rich to Ni-rich with longer annealing treatments, most likely due to the different diffusivities of Y and Ni in Al. Moreover, a spatial and chemical correlation between the nanorods and amorphous complexions is observed, suggesting that these complexions serve as preferential nucleation sites.

Precipitation plays an important role in controlling the properties of Al alloys, by restricting grain growth, increasing strength, and improving ductility [1,2,3]. Nucleation and growth of these second phase precipitates in commercial Al alloys is very important. For example, the plate-like $\eta'$ phase (MgZn$_2$) [4,5,6] can nucleate either homogeneously in the grain interior or heterogeneously near dislocation lines and junctions, depending on the matrix grain size [7].

When rare-earth elements are added into Al alloys, precipitates with a core-shell structure can form. For example, Clouet et al. [8] observed an inhomogeneous structure of precipitates with Sc-enriched cores and Zr-enriched shells in an aged Al-0.09Sc-0.03Zr (at.%) alloy. The core-shell structure was attributed to both the faster diffusivity of Sc compared to Zr in the solid solution and the absence of Zr and Sc diffusion within the precipitate. The core-shell chemistry can also evolve over time depending on differential diffusivies of the alloying elements, as Van Dalen et al. [9] reported that precipitates in an Al-0.06Sc-0.02Yb alloy were initially Yb-rich but then developed a Yb-rich core with an enveloping Sc-rich shell as more Sc atoms diffuse to the precipitates during progressive annealing.

Our previous study [10] uncovered abundant unique nanorod precipitates with a core-shell structure at grain boundaries of nanocrystalline Al alloys, with the core composed of Al and C and the shell being Y-enriched. The nanorods together with amorphous grain boundary complexions play a critical role in stabilizing the grain structure and strengthening the materials. Therefore, in the present study, a detailed characterization of the structural and chemical evolution of the nanorods in a nanocrystalline Al-2Ni-2Y (at.%) alloy during annealing is performed. The nanorods nucleate with a structurally disordered internal structure, and then some transform to an ordered, crystalline state and grow into adjacent grain interiors as their size increases. At the same time, the matrix-nanorod interface can be either coherent or incoherent. During the entire evolution process, the nanorod cores remain composed of Al and C, while the shell transitions from Y-rich to Ni-rich as annealing time increases. Moreover, a strong spatial correlation between the nanorods and amorphous complexions (see Ref. [11] for an introduction to these defect structures) is observed, suggesting that these complexions serve as the initial nucleation sites for nanorods.

The fabrication process for bulk nanocrystalline Al alloys was detailed in Ref. [10] and is therefore only briefly described here. First, powders of elemental Al, Ni, and Y were ball milled for 10 hours in a SPEX SamplePrep 8000M high-energy ball mill within a glovebox filled with Ar gas and with a $O_2$ level <0.05 ppm. Next, the powders were consolidated into bulk pellets using an MTI Corporation OTF-1200X-VHP3 hot press. The powders were first cold pressed for 10 min under 100 MPa at room temperature, and then hot pressed for 1 h under 100 MPa at 540 ºC, followed by naturally cooling down to room temperature. In addition to the as-consolidated pellet, two more pellets with the same fabrication process were subsequently annealed at 540 ºC (0.87 of the melting temperature of pure Al) for 600 s (10 min) and 86400 s (1 day). Following the annealing, each pellet was placed on an Al heat sink submerged in a large reservoir of liquid nitrogen, so that the surface containing the Al block was rapidly cooled to freeze in any high temperature microstructural features. All of the TEM studies were performed in the area very close to the rapidly cooled surface. Scanning electron microscopy combined with energy dispersive spectroscopy (EDS) was employed to examine the sample composition, which is Al-2.2 at.% Ni-2.1 at.% Y, in an FEI Quanta 3D FEG dual-beam SEM/Focused Ion Beam microscope. Conventional and scanning transmission electron microscopy ((S)TEM) paired with EDS was used to examine feature sizes/structure and local chemistry using a JEOL JEM-2800 S/TEM operated at 200 kV. High resolution high-angle annular dark-field (HAADF)-STEM and EDS were employed to study the structure and chemistry of both nanorods and amorphous complexions using a JEOL JEM-ARM300F Grand ARM TEM with double Cs correctors operated at 300 kV. For HAADF-STEM imaging, a probe current of 35 pA was used, and the inner and outer collection angles were 106 and 180 mrad, respectively. Hardness values for different annealing conditions were obtained by nanoindentation tests using a Nano Indenter G200 (Agilent Technologies) with a maximum penetration depth of 400 nm and a constant indentation strain rate of 0.05 s$^{-1}$.

Figure 1 displays the size evolution of both the face-centered cubic (FCC) grains and the nanorods (indicated by arrows), with Figure 1(a)-(c) presenting bright field (BF) TEM micrographs corresponding to different annealing times. With longer annealing time, the morphology of each feature remains self-similar – an equiaxed shape for the FCC grains and an elongated shape for nanorods – while the size increases. The nanorods always maintain some connectivity to a nearby grain boundary, with growth occurring into adjacent grains. To quantitatively study the size evolution, more than 100 Al-rich FCC grains and nanorods were measured, with the corresponding cumulative distribution functions of feature descriptors presented in Figure 1(d) and (e). For the nanorods, the length and width evolve from 14 nm and 3.6 nm to 149 nm and 25 nm, respectively, after annealing for 86400 s (1 day), while the FCC grain size increases from 44 nm to 347 nm during the same period. In order to compare the growth kinetics, Figure 1(f) plots feature size as a function of annealing time on a log-log scale because both grain growth and precipitate coarsening typically follow a power-law relationship with a relatively similar exponent [12,13,14]. The trend lines are only meant to guide the eye, yet the coarsening rates for the different features are generally very similar, suggesting that the features may either share similar growth mechanisms or be triggered by companion events [15,16,17]. For nanocrystalline materials, nanoscale secondary phases at the grain boundaries often lead to a Zener pinning effect, providing kinetic stabilization of the grain structure [18,19,20], which is likely to be the case for the nanorod precipitates [10]. Moreover, these nanorods significantly strengthen the materials. The hardness of the sample after annealing for 600 s was 2.1 GPa, roughly the same as the as-consolidated sample (2.0 GPa), even though the average grain size is ~2.5 times larger after annealing. Further annealing for 86400 s led to a decrease in hardness to 1.5 GPa.

The chemical evolution of the nanorods is studied using STEM-EDS in Figure 2. All annealing conditions show a core-shell nanorod structure, with the shell enriched with heavier elements. For $t_{ANN} = 0$ s (Figure 2(a)), the brighter shell is due to segregation of Y while the core is primarily Al and C. No obvious variation in Ni concentration occurs near the nanorods prior to annealing treatments. As $t_{ANN}$ increases to 600 s (Figure 2(b)), Ni tends to deplete from the nanorod interior region, while the distribution of Al, Y, and C is qualitatively the same as at $t_{ANN} = 0$ s. When $t_{ANN}$ reaches 86400 s (Figure 2(c)), Y is no longer localized to the shell only as some Y atoms also diffuse into the nanorod core, while Ni now more clearly segregates to the shell. Therefore, the nanorod shell transitions from Y-rich to Ni-rich with increasing annealing time and coarsening of the nanorod features.

Although precipitates with a core-shell structure have been observed in Al alloys, previous studies have reported invariant chemical compositions during annealing [8,9,21]. In contrast, we report here an evolution of the shell chemistry, which we hypothesize is due to the different diffusivities of Ni and Y in the Al solvent. Wu et al. [22] performed density functional theory calculations and showed that the diffusion barrier for Y in Al (~1.1 eV) is significantly smaller than that for Ni in Al (~1.6 eV). Consequently, Y diffuses faster than Ni and will reach the nanorod shell first. As annealing time increases, Y tends to react with O (Figure S1) so that there is less available Y to decorate the shell, and consequently Ni diffuses to these regions given sufficient time.  The segregation of Y or Ni atoms to the matrix-nanorod interface for all annealing conditions also suggests that the solute atoms must be dragged by the moving interface as the nanorods grow.  The composition of the nanorod core is possibly due to the energetically favorable phase, so the nanorod structure will be investigated next.

Figure 3 shows three different nanorod types observed for $t_{ANN} = 0$ s, as this state contained all structural variations and serves as a starting condition for describing evolution.  The first nanorod type has an incoherent interfacial structure and a disordered nanorod interior (Figure 3(a)), and are the smallest nanorods in the microstructure.  The brighter interfaces in the HAADF-STEM image correspond to an enrichment of Y atoms, as shown in Figure 2(a).  In addition, bright spots are randomly distributed within the matrix and nanorod, signifying Y incorporation in the nanorod interior and possible substitution of Al with Y atoms.  A magnified view of the enclosed area in the BF-STEM image shows an incoherent interface and a disordered nanorod interior, while the adjacent grains exhibit lattice fringes with different orientations (i.e., the nanorod is at a grain boundary).  Figure 3(b) displays the second nanorod type, where the interface remains incoherent but the nanorod interior is crystalline (i.e., ordered).  The interface is again bright in the HAADF-STEM micrograph, and one interfacial region is outlined in the BF-STEM micrograph.  The zoomed view reveals an incoherent interface ~1 nm thick, while the nanorod interior exhibits obvious lattice fringes.  Figure 3(c) shows the last nanorod type, which has a coherent interface and an ordered nanorod interior.  Two areas along the top and bottom interfaces are enclosed in the HAADF-STEM image, with the corresponding Fourier-filtered (FF) images showing a coherent interface and an ordered nanorod interior, where lattice connectivity is observed across the interface.  To further study the internal structure, Figure 3(d) presents a HAADF-STEM micrograph of a nanorod with its longer side parallel to the horizontal direction.  The interior structure is magnified in the second image and has a fast Fourier transform as an inset. Since Figure 2(a) shows that nanorod interior consists of Al and C, so the bright spots in the magnified view correspond to Al due to its larger atomic weight. To better visualize the structure, an FF image is presented in the third panel, and the atomic arrangement matches the schematic illustration of the $Al_4C_3$ phase (last panel). For the other two annealing times, all nanorods are found to have ordered nanorod interiors associated with this $Al_4C_3$ phase (Figure 3(e) and (f)), supporting the idea that nanorod growth to larger sizes coincides with crystallization of the features. The $Al_4C_3$ phase has been commonly used as a reinforcement in Al matrix composites with enhanced mechanical performance [23,24]. The carbide crystallizes in the R-3m space group with lattice parameters of $a$ = 0.33 nm and $c$ = 2.5 nm in the hexagonal setting [25,26,27]. A rod-like morphology is usually observed for this phase because of preferential growth along the <1 1 0> direction of $Al_4C_3$, due to the free energy of the prism plane surface being higher than that of the basal plane [28]. When dispersed in the matrix, the carbide phase could in theory have an orientation relationship with the Al matrix (e.g., Al [0 0 1]//$Al_4C_3$ [1 1 0] in Ref. [29]). In the present study however, the carbide nanorods do not have a specific orientation relationship with the Al matrix, possibly due to the nanorods nucleation occurring at amorphous grain boundary complexions rather than within the matrix. It should be noted that segregation of Y and /or Ni to the interface parallel to the longer side (the major axis) of the nanorod is stronger than to the shorter side (the minor axis), indicating a segregation tendency to the (0 0 1) plane. The effect may be due to the anisotropic interfacial energy of the $Al_4C_3$ phase [30].

The nanorods nucleate with a disordered internal structure, as the smallest nanorods possess such a structure and growth leads to crystallization, suggesting that the amorphous grain boundary complexions observed in these alloys might act as nucleation sites. Figure 4(a) presents low-magnification (LM) and high-magnification (HM) HAADF and BF-STEM micrographs of an amorphous complexion (outlined in yellow dashed lines) and connected nanorod for $t_{ANN} = 0$ s, supporting this hypothesis. The chemistry of a different amorphous complexion and its adjacent nanorod is shown in Figure 4(b). Elemental mapping reveals that both features are enriched with C and Y, while Ni atoms seem to segregate to only the matrix-complexion interface. For the nanorod, similar to Figure 2(a), the core and shell regions are enriched with C and Y atoms, respectively. Therefore, amorphous complexions and nanorods are both enriched with the same types of elements. It should be noted that for ordered FCC grain boundaries, only segregation of Ni and/or Y atoms was observed (Figure S2). The similar chemical and structural characteristics of the two feature types, as well as their spatial correlation (every nanorod investigated was connected to an amorphous complexion), strongly suggest that the complexions serve as nucleation sites and supplying dopant atoms to the nanorods.

In summary, the evolution of core-shell structured nanorods in a nanocrystalline Al-Ni-Y during annealing is studied. The nanorods nucleate with a disordered structure and transform to an ordered $Al_4C_3$ phase as they grow. A transition from a Y-rich to a Ni-rich shell is observed, while the core remains composed of Al and C for all annealing treatments. Finally, the nanorods are found to nucleate from amorphous complexions, which serve as a source of the solute atoms that feed the nanorod growth, and both features concurrently stabilize the grain structure as well as strengthen the material. These new insights shed light on a unique defect type in nanocrystalline Al alloys that can provide stabilization against coarsening and can improve strength.

**Declaration of Competing Interest**

The authors declare that they have no known competing financial interest or personal relationships that could have appeared to influence the work reported in this paper.

## Acknowledgements

This work was supported by the U.S. Department of Energy, Office of Energy Efficiency and Renewable Energy (EERE), under the Advanced Manufacturing Office Award No. DE-EE0009114.  The authors acknowledge the use of facilities and instrumentation at the UC Irvine Materials Research Institute (IMRI), which is supported in part by the National Science Foundation through the UC Irvine Materials Research Science and Engineering Center (DMR-2011967).  SEM, FIB, and EDS work was performed using instrumentation funded in part by the National Science Foundation Center for Chemistry at the Space-Time Limit (CHE-0802913).

# References

[1] S. Cheng, Y. H. Zhao, Y. T. Zhu, E. Ma, Acta Mater. 55 (2007) 5822-5832.

[2] K. Ma, H. Wen, T. Hu, T. D. Topping, D. Isheim, D. N. Seidman, E. J. Lavernia, J. M. Schoenung, Acta Mater. 62 (2014) 141-155.

[3] C. Y. Nam, J. H. Han, Y. H. Chung, M. C. Shin, Mater. Sci. Eng. A 347 (2003) 253-257.

[4] Y. H. Zhao, X. Z. Liao, Z. Jin, R. Z. Valiev, Y. T. Zhu, Acta Mater. 52 (2004) 4589-4599.

[5] K. R. Cardoso, D. N. Travessa, W. J. Botta, A. M. Jorge Jr., Mater. Sci. Eng. A 528 (2011) 5804-5811.

[6] Y. Reda, R. Abdel-Karim, I. Elmahallawi, Mater. Sci. Eng. A 485 (2008) 468-475.

[7] T. Hu, K. Ma, T. D. Topping, J. M. Schoenung, E. J. Lavernia, Acta Mater. 61 (2013) 2163-2178.

[8] E. Clouet, L. Lae, T. Epicier, W. Lefebvre, M. Nastar, A. Deschamps, Nat. Mater. 5 (2006) 482-488.

[9] M. E. Van Dalen, D. C. Dunand, D. N. Seidman, Acta Mater. 59 (2011) 5224-5237.

[10] T. Lei, J. Shin, D. S. Gianola, T. J. Rupert, Acta Mater. 221 (2021) 117394.

[11] J. Luo, Curr. Opin. Solid State Mater. Sci. 12 (2008) 81-88.

[12] H. V. Atkinson, Acta Metall. 36(3) (1988), 469-491.

[13] I. M. Lifshitz, V. V. Slyozov, J. Phys. Chem. Solids 19 (1-2) (1961) 35-50.

[14] C. Wagner, Z. Elektrochem 65 (1961) 581-591.

[15] G. Gottstein, Y. Ma, and L. S. Shvindlerman, Acta Mater. 53(5) (2005) 1535-1544.

[16] R. Dannenberg, E. Stach, J. R. Groza, and B. J. Dresser, Thin Solid Films 379 (1-2) (2000) 133-138.

[17] A. J. Ardell and V. Ozolins, Nat. Mater. 4 (2015) 309-316.

[18] C. C. Koch, R. O. Scattergood, M. Saber, H. Kotan, J. Mater. Res. 28(13) (2013) 1785-1791.

[19] S. Praveen, J. Basu, S. Kashyap, R. S. Kottada, J. Alloys Compd. 662 (2016) 361-367.

[20] Y. Z. Chen, K. Wang, G. B. Shan, A. V. Ceguerra, L. K. Huang, H. Dong, L. F. Cao, S. P. Ringer, F. Liu, Acta Mater. 158 (2018) 340-353.

[21] R. A. Karnesky, D. C. Dunand, D. N. Seidman, Acta Mater. 57 (2009) 4022-4031.

[22] H. Wu, T. Mayeshiba, D. Morgan, Sci. Data 3 (2016) 160054.

[23] B. Chen, L. Jia, S. Li, H. Imai, M. Takahashi, and K. Kondoh, Adv. Eng. Mater. 16(8) (2014) 972-975.

[24] A. Santos-Beltrán, R. Goytia-Reyes, H. Morales-Rodriguez, V. Gallegos-Orozco, M. Santos-Beltrán, F. Baldenebro-Lopez, and R. Martínez-Sánchez, Mater. Charact. 106 (2015) 368-374.

[25] A. Pisch, A. Pasturel, G. Deffrennes, O. Dezellus, P. Benigni, and G. Mikaelian, Comput. Mater. Sci. 171 (2020) 109100.

[26] T. M. Gesing and W. Jeitschko, Zeitschrift Für Naturforschung B 50(2) (1995) 196-200.

[27] A. Jain, S. P. Ong, G. Hautier, W. Chen, W. D. Richards, S. Dacek, S. Cholia, D. Gunter, D. Skinner, G. Ceder, and K. A. Persson, APL Mater. 1(1) (2013) 011002.

[28] L. Ci, Z. Ryu, N. Y. Jin-Phillipp, and M. Rühle, Acta Mater. 54 (2006) 5367-5375.

[29] W. Zhou, M. Dong, Z. Zhou, X. Sun, K. Kikuchi, N. Nomura, and A. Kawasaki, Carbon 141 (2019) 67-75.

[30] Z. Y. Zhao, W. J. Zhao, P. K. Bai, L. Y. Wu, P. C. Huo, Mater. Lett. 255 (2019) 126559.

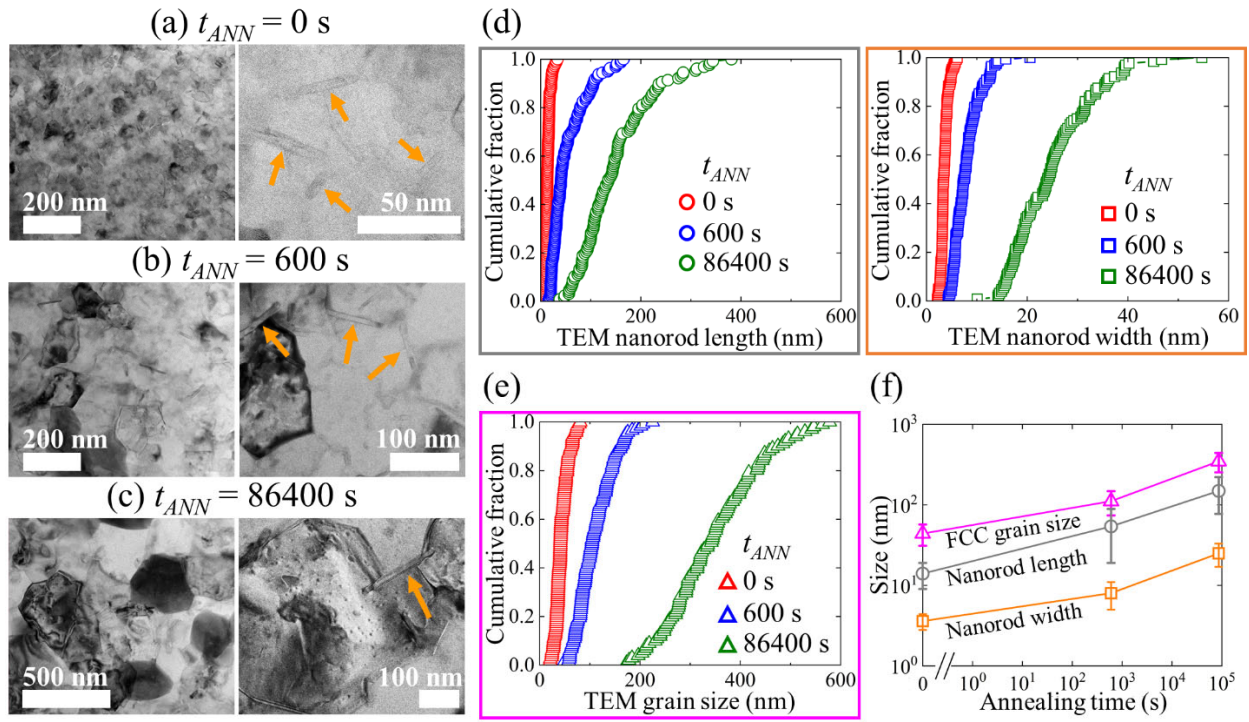

**Figure 1.** (a)-(c) Bright field TEM micrographs of the grain structure and nanorods (indicated by arrows) for three annealing times at a temperature of 540 °C. (d) Cumulative distribution functions of the measured nanorod lengths and widths. (e) Cumulative distribution functions of the Al-rich FCC grain size. (f) Average size as a function of annealing time for the FCC grains and the nanorod dimensions. The error bars represent the corresponding standard deviation.

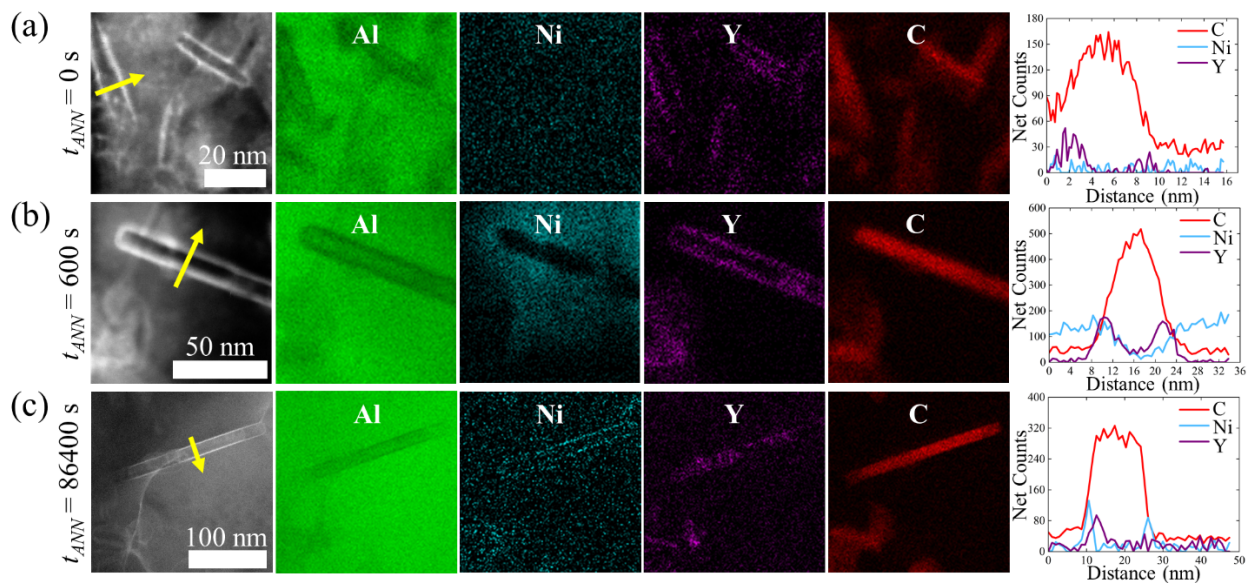

**Figure 2.** HAADF-STEM micrographs, the corresponding EDS elemental mapping, and line scans across nanorods (indicated by yellow arrows) for different annealing times, revealing the core-shell structure of the nanorods. The core of the nanorods remains composed of Al and C for all annealing times, while the outside shell transitions from Y-rich at (a) $t_{ANN}$ = 0 s and (b) $t_{ANN}$ = 600 s to Ni-rich after (c) $t_{ANN}$ = 86400 s, when some Y atoms diffuse into the interior of the nanorod.

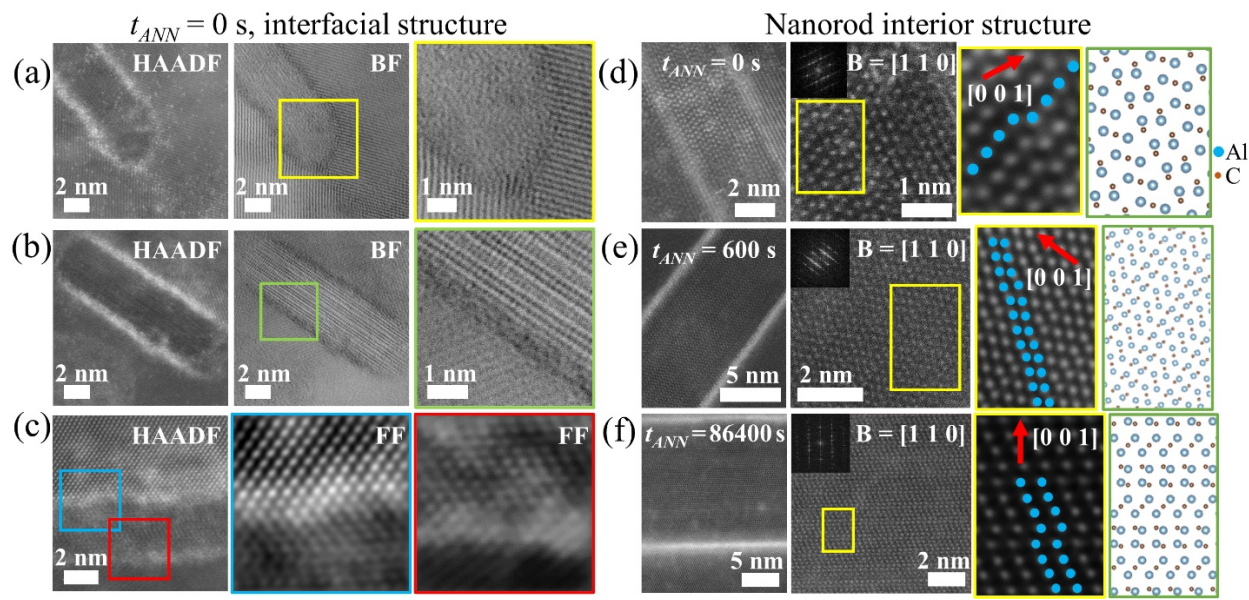

**Figure 3.** (a)-(c) High resolution STEM micrographs reveal three types of matrix-nanorod interfaces for $t_{ANN}$ = 0 s. (a) Incoherent interfaces can separate a disordered nanorod interior and the crystalline matrix. (b) Incoherent matrix-nanorod interfaces can also occur between an ordered nanorod interior and the surrounding matrix. (c) Coherent matrix-nanorod interfaces can exist between ordered nanorod interior and the matrix, where one upper interface (blue square) and one lower interface (red square) are presented as magnified Fourier-filtered (FF) image to show their local structure. (d)-(e) High resolution STEM micrographs show the ordered nanorod phase, $Al_4C_3$, for various annealing conditions – (d) $t_{ANN}$ = 0s, (e) $t_{ANN}$ = 600 s, $t_{ANN}$ = 86400 s. For each condition, the first micrograph shows the orientation of the nanorod, and the second one is a magnified view of the nanorod interior along with the corresponding fast Fourier transform. The third image is an FF image of the nanorod interior, and the last one exhibits a schematic illustration of the $Al_4C_3$ structure with the same orientation as that in the FF image.

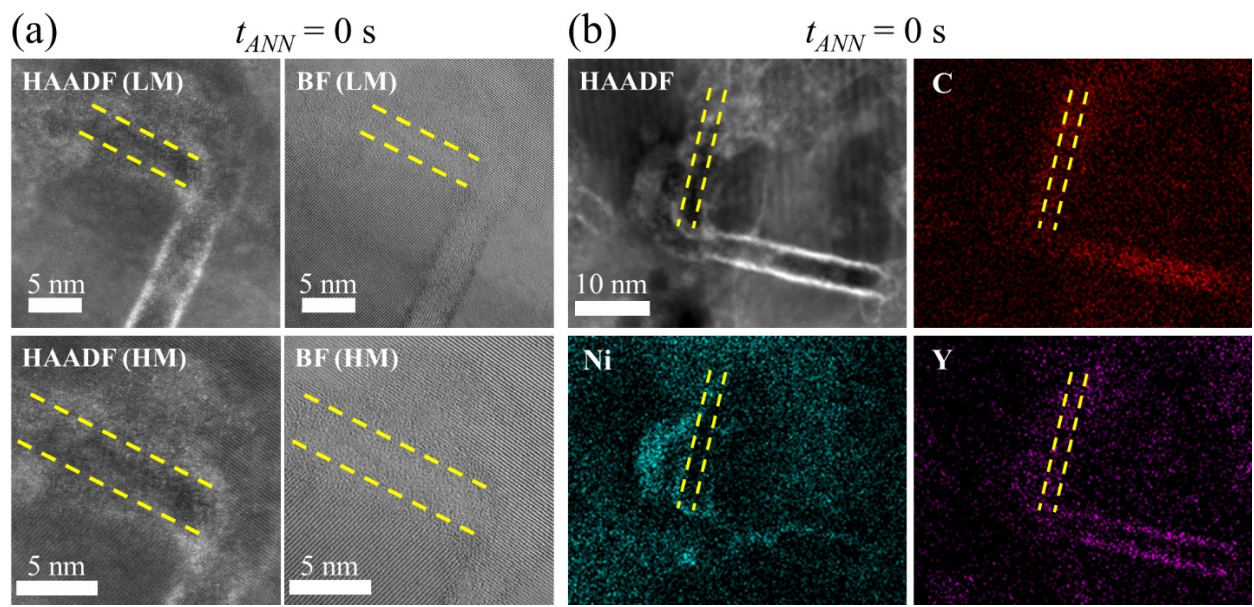

**Figure 4.** (a) HAADF-STEM and complementary BF-STEM micrographs of an amorphous complexion (outlined by dashed lines) close to a nanorod at low magnification (LM, top) and high magnification (HM, bottom) for $t_{ANN}$ = 0 s. (b) HAADF-STEM micrograph and EDS elemental mapping of one different amorphous complexion (outlined by dashed lines) and its adjacent nanorod reveals that both features are enriched in Y and C, while only the amorphous complexion region appear enriched in Ni.

**Supplementary Material**

$t_{ANN} = 86400$ s

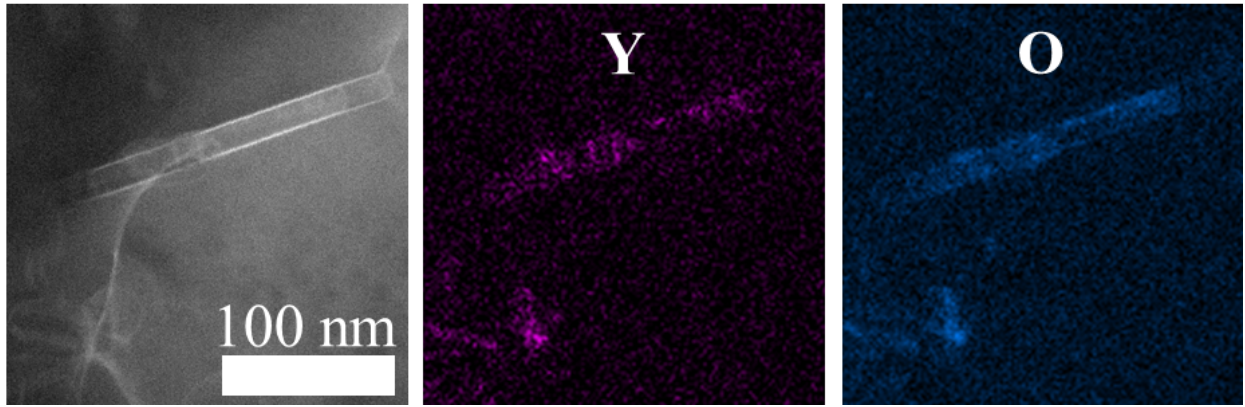

**Figure S1.** HAADF-STEM micrographs and the corresponding Y and O maps for $t_{ANN} = 86400$ s revealing that Y atoms tend to combine with O atoms for longer annealing time.

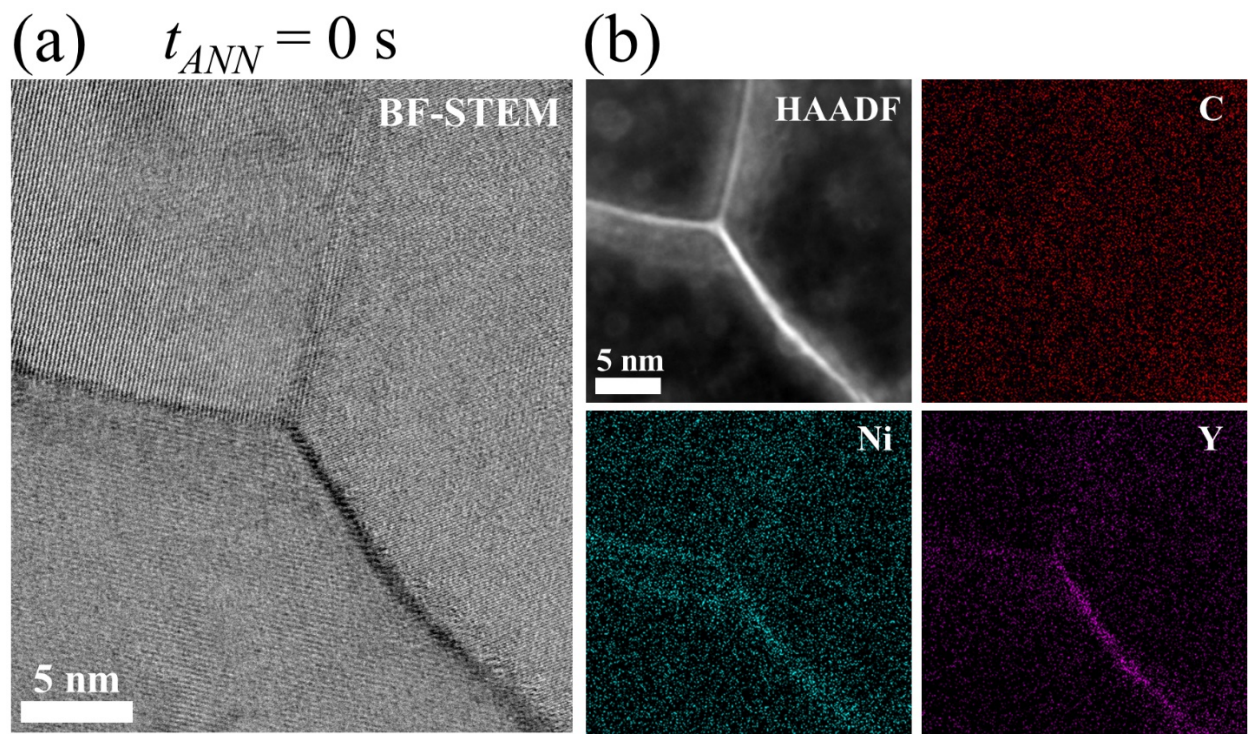

**Figure S2.** (a) High-resolution BF-STEM micrograph of three ordered grain boundaries for $t_{ANN}$ = 0 s. (b) HAADF-STEM micrograph and EDS elemental mapping show various segregation degrees of Ni and Y to the three boundaries. However, no segregation of C atoms is observed.